\documentstyle[12pt,prd,aps,epsfig,preprint]{revtex}

\begin{document}
\title{A perturbative treatment for the energy levels of neutral atoms }
\author{Sameer M. Ikhdair\thanks{%
sameer@neu.edu.tr} and \ Ramazan Sever\thanks{%
sever@metu.edu.tr}}
\address{$^{\ast }$Department of Physics, Near East University, Nicosia, North
Cyprus, Mersin-10, Turkey\\
$^{\dagger }$Department of Physics, Middle East Technical University, 06531
Ankara, Turkey.}
\maketitle

\begin{abstract}
Energy levels of neutral atoms have been re-examined by applying an
alternative perturbative scheme in solving the Schr\"{o}dinger equation for
the Yukawa potential model with a modified screening parameter. The
predicted shell binding energies are found to be quite accurate over the
entire range of the atomic number $Z$ up to 84 and compare very well with
those obtained within the framework of hyper-virial-Pad{\bf \'{e}} scheme
and the method of shifted large-N expansion. It is observed that the new
perturbative method may also be applied to the other areas of atomic
physics.\


PACS\ No: 03.65-W; 03.65.Ge; 03.65 Sq
\end{abstract}

\ \ \ \ \ \ \ \ \ \ \ \ \ \ \ \ \
\newpage
\section{Introduction}

\noindent In recent years the energy levels of neural atoms have been
studied by several analytic methods [1-8] in which it is assumed that the
screened potential of the atom may be of static screened Coulomb (SSC) which
is well represented by Yukawa form:

\begin{equation}
V\left( r\right) =-\left( \frac{A}{r}\right) \exp (-\delta r),
\end{equation}
with $A=\alpha Ze^{2},$ where $\alpha =(137.037)^{-1}$ is the fine-structure
constant and $Z$ is the atomic number. This form is often used for the
description of the energy levels of light to heavy neutral atoms [7]. It is
known that SSC potential yields reasonable results only for the innermost
states when $Z$ is large. However, for the outermost and middle atomic
states, it gives rather poor results. Although the bound state energies for
the SSC potential with $Z=1$ have also been studied. The screening parameter
$\delta $ is chosen to be

\begin{equation}
\delta =\delta _{0}Z^{1/3},
\end{equation}
corresponding to the $Z$-dependence of the reciprocal of the Thomas-Fermi
radius of the atom. However, these analytic works, perturbative as well as
nonperturbative, fail to yield accurate shell binding energies for light
atoms, particularly in the range $Z\leq 9.$ Subsequently, it has been
pointed out by Refs.[5,6] that the major source of errors perhaps lies in
the wrong choice of the $Z$-dependence of the screening parameter. Invoking
Fermi-Amaldi correction [9] in the context of Ecker-Weizel approximation
(EWA) method [10], Dutt and Varshni [5,6,7] have suggested a modified form

\begin{equation}
\delta =\delta _{0}Z^{1/3}\left( 1-1/Z\right) ^{2/3},
\end{equation}
with $\delta _{0}=0.98$. Clearly, when $Z=1$, $\delta $ vanishes and the
potential in (1) becomes the Coulomb potential as it should be. Correctness
of the choice of the modified screening parameter has been further justified
by the recent work of Lai and Madan [8]. They have shown \ that the
hypervirial-Pad{\bf \'{e}} scheme which failed to reproduce correct shell
binding energies for light atoms using the screening parameter given in (2)
[4], yields very accurate energy eigenvalues using the modified screening
coefficient in (3) [8]$.$ However, one of the short-comings of the
hypervirial-Pad{\bf \'{e}} technique is that it involves elaborate
computational time and effort for each numerical prediction. Lai and Madan
[8] have to consider upto eleven terms in the perturbation series for the
energy eigenvalues in order to ensure the convergence of the Pad{\bf \'{e}}
approximant $E(N,M)$. Furthermore, application of this method becomes quite
restricted due to nonavailability of compact analytic expressions for the
bound-state energies, eigenfunctions and normalization constants.

On the other hand, Dutt and Varshni [7] have investigated the bound-states
of neutral atoms using the large-$N$ expansion method which has been claimed
to be very powerful for solving potential problems in nonrelativistic
quantum mechanics. This technique also requires an approximate treatment and
computational time as well.

In this paper, we investigate the bound-state properties of SSC potential
using a new perturbative formalism [11] which has been claimed to be very
powerful for solving the Schr\"{o}dinger equation to obtain the bound-state
energies as well as the wave functions in Yukawa or SSC potential problem
[11,12] in both bound and continuum regions. This novel treatment is based
on the decomposition of the radial Schr\"{o}dinger equation into two pieces
having an exactly solvable part with an addi\i tional piece leading to
either a closed analytical solution or approximate treatment depending on
the nature of the perturbed potential.

It seems then logical and meaningful to probe whether the range of
applicability of this novel perturbation treatment may be widened. As a
first attempt, we have shown recently \ that the method adequately explains
the spectrum of hydrogen-like atoms $A=Z=1$ and also light and heavy atoms
[13]. With a view to make further applications to problems of atomic
physics, we compute here the shell binding energies of light to heavy
neutral atoms. The relevant steps of the perturbation scheme are to obtain
analytical expressions for the bound-state energy levels and corresponding
normalized eignfunctions.

The contents of this paper is as follows. In section \ref{TM} we breifly
outline the method with all necessary formulae to perform the current
calculations. In section \ref{A} we apply the approach to the
Schr\"{o}dinger equation with SSC potential and present the results obtained
analytically and numerically for the bound-state energy values upto third
perturbation energy shift. Finally, in section \ref{CR} we give our
concluding remarks.

\section{The Method}

\label{TM}For the consideration of spherically symmetric potentials, the
corresponding Schr\"{o}dinger equation, in the bound state domain, for the
radial wave function reads

\begin{equation}
\frac{\hbar ^{2}}{2m}\frac{\psi _{n}^{\prime \prime }(r)}{\psi _{n}\left(
r\right) }=V(r)-E_{n},
\end{equation}
with

\begin{equation}
V\left( r\right) =\left[ V_{0}(r)+\frac{\hbar ^{2}}{2m}\frac{\ell (\ell +1)}{%
r^{2}}\right] +\Delta V(r),
\end{equation}
where $\Delta V(r)$ is a perturbing potential and $\psi _{n}(r)=\chi
_{n}(r)u_{n}(r)$ is the full radial wave function, in which $\chi _{n}(r)$
is the known normalized eigenfunction of the unperturbed Schr\"{o}dinger
equation whereas $u_{n}(r)$ is a moderating wave function corresponding to
the perturbing potential. Following the prescription of Refs. [11,12], we
may rewrite (4) in the form:

\begin{equation}
\frac{\hbar ^{2}}{2m}\left( \frac{\chi _{n}^{\prime \prime }(r)}{\chi _{n}(r)%
}+\frac{u_{n}^{\prime \prime }(r)}{u_{n}(r)}+2\frac{\chi _{n}^{\prime
}(r)u_{n}^{\prime }(r)}{\chi _{n}(r)u_{n}(r)}\right) =V(r)-E_{n}.
\end{equation}
The logarithmic derivatives of the unperturbed $\chi _{n}(r)$ and perturbed $%
u_{n}(r)$ wave functions are given by

\begin{equation}
W_{n}(r)=-\frac{\hbar }{\sqrt{2m}}\frac{\chi _{n}^{\prime }(r)}{\chi _{n}(r)}%
\text{ \ \ and \ \ }\Delta W_{n}=-\frac{\hbar }{\sqrt{2m}}\frac{%
u_{n}^{\prime }(r)}{u_{n}(r)},
\end{equation}
which leads to

\begin{equation}
\frac{\hbar ^{2}}{2m}\frac{\chi _{n}^{\prime \prime }(r)}{\chi _{n}(r)}%
=W_{n}^{2}(r)-\frac{\hbar }{\sqrt{2m}}W_{n}^{^{\prime }}(r)=\left[ V_{0}(r)+%
\frac{\hbar ^{2}}{2m}\frac{\ell (\ell +1)}{r^{2}}\right] -\varepsilon _{n},
\end{equation}
where $\varepsilon _{n}$ is the eigenvalue for the exactly solvable
potential of interest, and

\begin{equation}
\frac{\hbar ^{2}}{2m}\left( \frac{u_{n}^{\prime \prime }(r)}{u_{n}(r)}+2%
\frac{\chi _{n}^{\prime }(r)u_{n}^{\prime }(r)}{\chi _{n}(r)u_{n}(r)}\right)
=\Delta W_{n}^{2}(r)-\frac{\hbar }{\sqrt{2m}}\Delta W_{n}^{\prime
}(r)+2W_{n}(r)\Delta W_{n}(r)=\Delta V(r)-\Delta \varepsilon _{n},
\end{equation}
in which $\Delta \varepsilon _{n}=E_{n}^{(1)}+E_{n}^{(2)}+E_{n}^{(3)}+\cdots
$ is the correction term to the energy due to $\Delta V(r)$ and $%
E_{n}=\varepsilon _{n}+\Delta \varepsilon _{n}.$ If Eq. (9), which is the
most significant piece of the present formalism, can be solved analytically
as in (8), then the whole problem, in Eq. (4) reduces to the following form

\begin{equation}
\left[ W_{n}(r)+\Delta W_{n}(r)\right] ^{2}-\frac{\hbar }{\sqrt{2m}}\left[
W_{n}(r)+\Delta W_{n}(r)\right] ^{\prime }=V(r)-E_{n},
\end{equation}
which is a well known treatment within the frame of supersymmetric quantum
theory (SSQT) [14]. Thus, if the whole spectrum and corresponding
eigenfunctions of the unperturbed interaction potential are known, then one
can easily calculate the required superpotential $W_{n}(r)$ for any state of
interest leading to direct computation of related corrections to the
unperturbed energy and wave function.

For the perturbation technique, we can split the given potential in Eq.(4)
into two parts. The main part corresponds to a shape invariant potential,
Eq. (8), for which the superpotential is known analytically and the
remaining part is treated as a perturbation, Eq. (9). Therefore, it is
obvious that SSC potential can be treated using this prescription. In this
regard, the zeroth-order term corresponds to the Coulomb potential while
higher-order terms consitute the perturbation. However, the perturbation
term in its present form cannot be solved exactly through Eq. (9). Thus, one
should expand the functions related to the perturbation in terms of the
perturbation parameter $\lambda $,

\begin{equation}
\Delta V(r;\lambda )=\sum_{i=1}^{\infty }\lambda _{i}V_{i}(r),\text{ \ \ }%
\Delta W_{n}(r;\lambda )=\sum_{i=1}^{\infty }\lambda _{i}W_{n}^{(i)}(r),%
\text{ \ }E_{n}^{(i)}(\lambda )=\sum_{i=1}^{\infty }\lambda _{i}E_{n}^{(i)},
\end{equation}
where $i$ denotes the perturbation order. Substitution of the above
expansions into Eq. (9) and equating terms with the same power of $\lambda $%
\ on both sides up to $O(\lambda ^{4})$ gives

\begin{equation}
2W_{n}(r)W_{n}^{(1)}(r)-\frac{\hbar }{\sqrt{2m}}\frac{dW_{n}^{(1)}(r)}{dr}%
=V_{1}(r)-E_{n}^{(1)},
\end{equation}

\begin{equation}
W_{n}^{(1)}(r)W_{n}^{(1)}(r)+2W_{n}(r)W_{n}^{(2)}(r)-\frac{\hbar }{\sqrt{2m}}%
\frac{dW_{n}^{(2)}(r)}{dr}=V_{2}(r)-E_{n}^{(2)},
\end{equation}

\begin{equation}
2\left[ W_{n}(r)W_{n}^{(3)}(r)+W_{n}^{(1)}(r)W_{n}^{(2)}(r)\right] -\frac{%
\hbar }{\sqrt{2m}}\frac{dW_{n}^{(3)}(r)}{dr}=V_{3}(r)-E_{n}^{(3)},
\end{equation}
\begin{equation}
2\left[ W_{n}(r)W_{n}^{(4)}(r)+W_{n}^{(1)}(r)W_{n}^{(3)}(r)\right]
+W_{n}^{(2)}(r)W_{n}^{(2)}(r)-\frac{\hbar }{\sqrt{2m}}\frac{dW_{n}^{(4)}(r)}{%
dr}=V_{4}(r)-E_{n}^{(4)}.
\end{equation}
Hence, unlike the other perturbation theories, Eq. (9) and its expansion,
Eqs. (12-15), give a flexibility for the easy calculations of the
perturbative corrections to energy and wave functions for the $nth$ state of
interest through an appropriately chosen perturbed superpotential.

\section{Application to the SSC Potential}

\label{A}Considering the recent interest in various power-law potentials in
the literature, we work through the article within the frame of low
screening parameter. In this case, the SSC or Yukawa potential can be
expanded in power series of the screening parameter $\delta $ as [15]

\begin{equation}
V(r)=-\left( \frac{A}{r}\right) \exp (-\delta r)=-\left( \frac{A}{r}\right)
\sum_{i=0}^{\infty }V_{i}(\delta r)^{i},
\end{equation}
where the perturbation coefficients $V_{i}$ are given by

\begin{equation}
V_{1}=-1,\text{ }V_{2}=1/2,\text{ }V_{3}=-1/6,\text{ }V_{4}=1/24,\text{ }%
\cdots .
\end{equation}
We now apply this approximation method to Yukawa potential with the angular
momentum barrier

\begin{equation}
V(r)=-\left( \frac{A}{r}\right) \exp (-\delta r)+\frac{\ell (\ell +1)\hbar
^{2}}{2mr^{2}}=\left[ V_{0}(r)+\frac{\ell (\ell +1)\hbar ^{2}}{2mr^{2}}%
\right] +\Delta V(r),
\end{equation}
where the first piece is the shape invariant zeroth-order which is an
exactly solvable piece corresponding to the unperturbed Coulomb potential
with $V_{0}(r)=-A/r$ while $\Delta V(r)=A\delta -(A\delta ^{2}/2)r+(A\delta
^{3}/6)r^{2}-(A\delta ^{4}/24)r^{3}+\cdots $ is the perturbation term. The
literature is rich with examples of particular solutions for such power-law
potentials employed in different fields of physics, for recent applications
see Refs. [16,17]. At this stage one may wonder why the series expansion is
truncated at a lower order. This can be understood as follows. It is widely
appreciated that convergence is not an important or even desirable property
for series approximations in physical problems. Specifically, a slowly
convergent approximation which requires many terms to achieve reasonable
accuracy is much less valuable than the divergent series which gives
accurate answers in a few terms. This is clearly the case for the Yukawa
problem [18]. However, it is worthwhile to note that the main contributions
come from the first four terms. Thereby, the present calculations are
performed upto the third-order involving only these additional potential
terms, which suprisingly provide highly accurate results for small screening
parameter $\delta .$

\subsection{Ground State Calculations $\left( n=0\right) $}

In the light of Eq. (8), the zeroth-order calculations leading to exact
solutions can be carried out readily by setting the ground-state
superpotential and the unperturbed exact energy as

\begin{equation}
W_{n=0}\left( r\right) =-\frac{\hbar }{\sqrt{2m}}\ \frac{\ell +1}{r}+\sqrt{%
\frac{m}{2}}\frac{A}{(\ell +1)\hbar },\text{ \ \ }E_{n}^{(0)}=-\frac{mA^{2}}{%
2\hbar ^{2}(n+\ell +1)^{2}},\text{ \ \ \ }n=0,1,2,....
\end{equation}
and from the literature, the corresponding normalized Coulomb bound-state
wave function [19]

\begin{equation}
\chi _{n}^{(C)}(r)=N_{n,l}^{(C)}r^{\ell +1}\exp \left[ -\beta r\right]
\times L_{n}^{2\ell +1}\left[ 2\beta r\right] ,
\end{equation}
in which $N_{n,l}^{(C)}=\left[ \frac{2mA}{\left( n+\ell +1\right) \hbar ^{2}}%
\right] ^{\ell +1}\frac{1}{(n+\ell +1)}\frac{1}{\sqrt{\frac{\hbar ^{2}}{mAn!}%
(n+2\ell +1)!}}$ is a normalized constant,\ \ $\beta =\frac{mA}{\left(
n+\ell +1\right) \hbar ^{2}}$ and $L_{n}^{k}\left( x\right)
=\sum_{m=0}^{n}(-1)^{m}\frac{(n+k)!}{\left( n-m\right) !(m+k)!m!}x^{m}$ is
an associate Laguarre polynomial function [20].

For the sake of calculation of corrections to the zeroth-order energy and
wavefunction, one needs to consider the expressions leading to the first-
and third-order perturbation given by Eqs. (12--15). Multiplication of each
term in these equations by $\chi _{n}^{2}(r)$, and bearing in mind the
superpotentials given in Eq. (7), one can obtain the straightforward
expressions for the first-order correction to the energy and its
superpotential:
\begin{equation}
E_{n}^{(1)}=\int_{-\infty }^{\infty }\chi _{n}^{2}(r)\left( -\frac{A\delta
^{2}}{2}r\right) dr,\text{ }W_{n}^{(1)}\left( r\right) =\frac{\sqrt{2m}}{%
\hbar }\frac{1}{^{X_{n}^{2}(r)}}\int^{r}\chi _{n}^{2}(x)\left[ E_{n}^{(1)}+%
\frac{A\delta ^{2}}{2}x\right] dx,\
\end{equation}
and for the second-order correction and its superpotential:

\[
E_{n}^{(2)}=\int_{-\infty }^{\infty }\chi _{n}^{2}(r)\left[ \frac{A\delta
^{3}}{6}r^{2}-W_{n}^{(1)}\left( r\right) W_{n}^{(1)}\left( r\right) \right]
dr,\text{ }
\]
\begin{equation}
W_{n}^{(2)}\left( r\right) =\frac{\sqrt{2m}}{\hbar }\frac{1}{^{X_{n}^{2}(r)}}%
\int^{r}\chi _{n}^{2}(x)\left[ E_{n}^{(2)}+W_{n}^{(1)}\left( x\right)
W_{n}^{(1)}(x)-\frac{A\delta ^{3}}{6}x^{2}\right] dx,
\end{equation}
and for the third-order correction and its superpotential:

\[
E_{n}^{(3)}=\int_{-\infty }^{\infty }\chi _{n}^{2}(r)\left[ -\frac{A\delta
^{4}}{24}r^{3}-W_{n}^{(1)}\left( r\right) W_{n}^{(2)}\left( r\right) \right]
dr,\text{ }
\]
\begin{equation}
W_{n}^{(3)}\left( r\right) =\frac{\sqrt{2m}}{\hbar }\frac{1}{^{X_{n}^{2}(r)}}%
\int^{r}\chi _{n}^{2}(x)\left[ E_{n}^{(3)}+W_{n}^{(1)}(x)W_{n}^{(2)}(x)+%
\frac{A\delta ^{4}}{24}x^{3}\right] dx,
\end{equation}
for any state of interest. The above expressions calculate $W_{n}^{(1)}(r),$
$W_{n}^{(2)}(r)$\ and $W_{n}^{(3)}(r)$ explicitly from the energy
corrections $E_{n}^{(1)},$ $E_{n}^{(2)}$ and $E_{n}^{(3)}$ respectively,
which are in turn used to calculate the moderating wave function $u_{n}(r).$

Thus, through the use of Eqs. (21-23), one finds the ground state energy
shift upto the third-order and their moderating superpotentials as

\[
E_{0}^{(1)}\ =-\frac{\hbar ^{2}(3N_{0}^{2}-L)}{4m}\delta ^{2},
\]

\[
E_{0}^{(2)}=\frac{\hbar ^{4}N_{0}^{2}\left( 5N_{0}^{2}-3L+1\right) }{12Am^{2}%
}\delta ^{3}-\frac{\hbar ^{6}N_{0}^{4}\left( 5N_{0}^{2}-3L+1\right) }{%
16A^{2}m^{3}}\delta ^{4},
\]

\begin{eqnarray*}
E_{0}^{(3)}\ &=&-\frac{\hbar ^{6}N_{0}^{2}\left( 5N_{0}^{2}-3L\right) \left(
5N_{0}^{2}-3L+1\right) }{96A^{2}m^{3}}\delta ^{4}+\frac{\hbar
^{8}N_{0}^{4}\left( 5N_{0}^{2}-3L+1\right) \left( 9N_{0}^{2}-5L\right) }{%
48A^{3}m^{4}}\delta ^{5} \\
&&-\frac{\hbar ^{10}N_{0}^{6}\left( 5N_{0}^{2}-3L+1\right) \left(
9N_{0}^{2}-5L\right) }{64A^{4}m^{5}}\delta ^{6},
\end{eqnarray*}

\[
W_{0}^{(1)}(r)=-\frac{\hbar N_{0}\delta ^{2}}{2\sqrt{2m}}r,
\]
\ \ \ \ \ \ \ \ \ \ \ \ \ \ \ \ \ \ \ \ \ \ \ \ \ \ \ \ \ \ \ \ \ \ \ \ \ \
\ \ \ \ \ \ \ \ \ \ \ \ \ \ \ \ \ \ \ \ \ \ \ \ \ \ \ \ \ \ \ \ \ \ \ \ \ \
\ \ \ \ \ \ \ \ \ \ \ \ \ \ \ \ \ \ \ \ \ \ \ \ \ \ \ \ \ \ \ \ \ \ \ \ \ \
\ \
\begin{equation}
W_{0}^{(2)}(r)=-\frac{\hbar N_{0}\left[ Amr+\hbar ^{2}N_{0}N_{1}\right] %
\left[ 3\hbar ^{2}N_{0}^{2}\delta -4mA\right] \delta ^{3}}{24\sqrt{2m}%
(Am)^{2}}r,
\end{equation}
where $N_{0}=\left( \ell +1\right) ,$ $N_{1}=\left( \ell +2\right) $ and $%
L=\ell (\ell +1).$ Therefore, the analytical expressions for the lowest
energy and full radial wave function of the Yukawa potential are then given
by

\begin{equation}
E_{n=0,\ell }=E_{n=0}^{(0)}+A\delta
+E_{0}^{(1)}+E_{0}^{(2)}+E_{0}^{(3)}+\cdots ,\text{ }\psi _{n=0,\ell
}(r)\approx \chi _{n=0,\ell }^{(C)}(r)u_{n=0,\ell }(r),
\end{equation}
in which

\begin{equation}
u_{n=0,\ell }(r)\approx \exp \left( -\frac{\sqrt{2m}}{\hbar }\int^{r}\left(
W_{0}^{(1)}\left( x\right) +W_{0}^{(2)}\left( x\right) \right) dx\right) .
\end{equation}
Hence, the explicit form of the full wave function in (25) for the ground
state is

\begin{equation}
\psi _{n=0,\ell }(r)=\left[ \frac{2mA}{(\ell +1)\hbar ^{2}}\right] ^{\ell +1}%
\frac{1}{(\ell +1)^{2}}\sqrt{\frac{Am}{\hbar ^{2}(2\ell +1)!}}r^{\ell
+1}\exp (P(r)),
\end{equation}
with $P(r)=\sum_{i=1}^{5}p_{i}r^{i}$ is a polynomial of fifth order having
the following coefficients:
\begin{equation}
p_{1}=\frac{(\ell +1)}{A}E_{0}^{(2)}-\frac{Am}{(\ell +1)\hbar ^{2}},\text{ \
}p_{2}=\frac{9}{4}\frac{(\ell +2)}{(\ell +1)^{2}}c^{2}d\delta ^{4},\text{ \ }%
p_{3}=\frac{1}{6}cd\delta ^{4},\text{ }p_{4}=\frac{1}{8}ac\delta ^{4},\text{
}p_{5}=\frac{1}{10}c\delta ^{6},\text{\ }
\end{equation}
in which $d=b+\frac{6Am}{\hbar ^{2}(\ell +1)^{2}\delta }$ and other
parameters are given in (21).

\subsection{Excited state calculations $(n\geq 1)$}

The calculations procedures lead to a handy recursion relations in the case
of ground states, but becomes extremely cumbersome in the description of
radial excitations when nodes of wavefunctions are taken into account, in
particular during the higher order calculations. Although several attempts
have been made to bypass this difficulty and improve calculations in dealing
with excited states, (cf. e.g. [21], and the references therein) within the
frame of supersymmetric quantum mechanics (SUSYQM).

Using Eqs. (7) and (19), the superpotential $W_{n}(r)$ which is related to
the excited states can be readily calculated through Eqs. (21-23). So the
first-order energy shift in the first excited state $(n=1)$ and its
superpotential are

\[
E_{1}^{(1)}=-\frac{\hbar ^{2}(3N_{1}^{2}-L)}{4m}\delta ^{2},
\]
\

\begin{equation}
W_{1}^{(1)}(r)\approx -\frac{\hbar N_{1}\delta ^{2}}{2\sqrt{2m}}r.
\end{equation}
Consequently, the use of the approximated $W_{1}^{(1)}(r)$ in the preceeding
equation in (22) gives the energy correction in the second-order as\

\begin{equation}
\ E_{1}^{(2)}=\frac{\hbar ^{4}N_{1}^{2}\left( 5N_{1}^{2}-3L+1\right) }{%
12Am^{2}}\delta ^{3}-\frac{\hbar ^{6}N_{1}^{4}\left( 5N_{1}^{2}-3L+1\right)
}{16A^{2}m^{3}}\delta ^{4},
\end{equation}
We also find its supersymmetric potential

\begin{equation}
W_{1}^{(2)}(r)=-\frac{\hbar N_{1}\left[ Amr+\hbar ^{2}N_{1}N_{2}\right] %
\left[ 3\hbar ^{2}N_{1}^{2}\delta -4mA\right] \delta ^{3}}{24\sqrt{2m}%
(Am)^{2}}r,
\end{equation}
which gives the energy shift in the third-order as

\begin{eqnarray}
E_{1}^{(3)}\ &=&-\frac{\hbar ^{6}N_{1}^{2}\left( 5N_{1}^{2}-3L\right) \left(
5N_{1}^{2}-3L+1\right) }{96A^{2}m^{3}}\delta ^{4}+\frac{\hbar
^{8}N_{1}^{4}\left( 5N_{1}^{2}-3L+1\right) \left( 9N_{1}^{2}-5L\right) }{%
48A^{3}m^{4}}\delta ^{5} \\
&&-\frac{\hbar ^{10}N_{1}^{6}\left( 5N_{1}^{2}-3L+1\right) \left(
9N_{1}^{2}-5L\right) }{64A^{4}m^{5}}\delta ^{6},  \nonumber
\end{eqnarray}
Therefore, the approximated energy value of the Yukawa potential
corresponding to the first excited state is\

\begin{equation}
E_{n=1,\ell }=E_{1}^{(0)}+A\delta
+E_{1}^{(1)}+E_{1}^{(2)}+E_{1}^{(3)}+\cdots .
\end{equation}

The related radial wavefunction can be expressed in an analytical form in
the light of Eqs (21-23) and Eq.(25), if required. The appromation used in
this work would not affect considerably the sensitivity of the calculations.
On the other hand, it is found analytically that our investigations put
forward an interesting hierarchy between $W_{n}^{(1)}(r)$ terms of different
quantum states in the first order after circumventing the nodal difficulties
elegantly,\ \ \

\begin{equation}
W_{n}^{(1)}(r)\approx -\frac{\hbar \left( n+\ell +1\right) \delta ^{2}}{2%
\sqrt{2m}}r,
\end{equation}
which, for instance, for the second excited state $\left( n=2\right) $ leads
to the first-order correction

\[
\ E_{2}^{(1)}=-\frac{\hbar ^{2}(3N_{2}^{2}-L)}{4m}\delta ^{2},
\]

\begin{equation}
W_{2}^{(1)}(r)\approx -\frac{\hbar N_{2}\delta ^{2}}{2\sqrt{2m}}r.
\end{equation}
Thus, the use of the approximated $W_{2}^{(1)}(r)$ in the preceeding
equation (22) gives the energy shift in the second-order and its
superpotential as\

\[
\ E_{2}^{(2)}=\frac{\hbar ^{4}N_{2}^{2}\left( 5N_{2}^{2}-3L+1\right) }{%
12Am^{2}}\delta ^{3}-\frac{\hbar ^{6}N_{2}^{4}\left( 5N_{2}^{2}-3L+1\right)
}{16A^{2}m^{3}}\delta ^{4},.
\]
\begin{equation}
W_{2}^{(2)}(r)=-\frac{\hbar N_{2}\left[ Amr+\hbar ^{2}N_{2}N_{3}\right] %
\left[ 3\hbar ^{2}N_{2}^{2}\delta -4mA\right] \delta ^{3}}{24\sqrt{2m}%
(Am)^{2}}r,
\end{equation}
which leads, via Eq.(23), into the third-order energy shift

\begin{eqnarray}
E_{2}^{(3)}\ &=&-\frac{\hbar ^{6}N_{2}^{2}\left( 5N_{2}^{2}-3L\right) \left(
5N_{2}^{2}-3L+1\right) }{96A^{2}m^{3}}\delta ^{4}+\frac{\hbar
^{8}N_{2}^{4}\left( 5N_{2}^{2}-3L+1\right) \left( 9N_{2}^{2}-5L\right) }{%
48A^{3}m^{4}}\delta ^{5} \\
&&-\frac{\hbar ^{10}N_{2}^{6}\left( 5N_{2}^{2}-3L+1\right) \left(
9N_{2}^{2}-5L\right) }{64A^{4}m^{5}}\delta ^{6},  \nonumber
\end{eqnarray}
where $N_{2}=(\ell +3).$ Therefore, the approximated energy eigenvalue of
the Yukawa potential corresponding to the second excited state $(n=2)$ is\

\begin{equation}
E_{n=2,\ell }=E_{2}^{(0)}+A\delta +E_{2}^{(1)}+E_{2}^{(2)}+E_{2}^{(3)}\cdots
.
\end{equation}
Finally, from the supersymmetry, we find out the $nth-$ state energy shifts
together with their supersymmetric potentials as

\[
E_{n}^{(1)}\ =-\frac{\hbar ^{2}\left[ 3(n+l+1)^{2}-L\right] }{4m}\delta
^{2},
\]

\[
W_{n}^{(1)}(r)\approx -\frac{\hbar (n+l+1)\delta ^{2}}{2\sqrt{2m}}r.
\]

\[
E_{n}^{(2)}=\frac{\hbar ^{4}(n+l+1)^{2}\left[ 5(n+l+1)^{2}-3L+1\right] }{%
12Am^{2}}\delta ^{3}-\frac{\hbar ^{6}(n+l+1)^{4}\left[ 5(n+l+1)^{2}-3L+1%
\right] }{16A^{2}m^{3}}\delta ^{4},
\]

\[
W_{n}^{(2)}(r)=-\frac{\hbar ^{4}(n+l+1)\left[ Amr+\hbar ^{2}(n+l+1)(n+l+2)%
\right] \left[ 3\hbar ^{2}(n+l+1)^{2}\delta -4mA\right] \delta ^{3}}{24\sqrt{%
2m}(Am)^{2}}r,
\]

\begin{eqnarray}
E_{n}^{(3)}\ &=&-\frac{\hbar ^{6}(n+l+1)^{2}\left[ 5(n+l+1)^{2}-3L\right] %
\left[ 5(n+l+1)^{2}-3L+1\right] }{96A^{2}m^{3}}\delta ^{4} \\
&&+\frac{\hbar ^{8}(n+l+1)^{4}\left[ 5(n+l+1)^{2}-3L+1\right] \left[
9(n+l+1)^{2}-5L\right] }{48A^{3}m^{4}}\delta ^{5}  \nonumber \\
&&-\frac{\hbar ^{10}(n+l+1)^{6}\left( 5(n+l+1)^{2}-3L+1\right) \left(
9(n+l+1)^{2}-5L\right) }{64A^{4}m^{5}}\delta ^{6},  \nonumber
\end{eqnarray}
Thus, the total energy for the $nth-$state is

\begin{equation}
E_{n,\ell }=E_{n}^{(0)}+A\delta +E_{n}^{(1)}+E_{n}^{(2)}+E_{n}^{(3)}+\cdots .
\end{equation}
For the numerical results, in Tables 1--4, we list our calculated $K$ and $L$%
-shell binding energies for some values of $Z$ and compare those with the
hypervirial-Pad{\bf \'{e}} results [8], the shifted large-$N$ expansion
method [7] and the experimental values [23] for the $s$-state energies $%
E_{00}$ and $E_{10}$ and also the $p$-state energies $E_{01}$ and $E_{11}.$
It is observed that inspite of calculational simplicity, the present
approach yields results as accurate as predicted by more elaborate
hypervirial-Pad{\bf \'{e}} and shifted large-$N$ expansion calculation.
Finally, the scope of extending the method to calculate oscillator strength,
bound-bound transition matrix elements etc. which have significant
importance in atomic physics is also possible.

\section{Concluding Remarks}

\label{CR} Table 1-3 we present our calculated $K-$shall binding energies $%
E_{00}$ and $E_{01}$ and $L-$shell binding energies $E_{10}$ and $E_{11}$
for some values of $Z$ and compare them with the predictions of Lai and
Madan [8] and the experimental values [17]. We quote only the Pade
approximant $E$ [10,11] results which provide upper bound to the energy
eigenvalues. For $E_{10\text{ }}$and $E_{20}$ levels. We also depict our
earlier results [5] obtained through EWA method which provides compact
analytic expressions only for the bound $s$-state energy eigenvalues. As we
have used through out the atomic units, our energies are measured in units
of 2$Ry=27.212$ $eV$ $\left[ 18\right] $ \ is used .One may notice that in
comparison to our earlier calculation based on EWA method, the present
techniques gives much improved energy eigenvalues. Furthermore, our
predictions are surprisingly close to those obtained through the use of
elaborate hypervirial technique. This indicates that there is distinct
advantage in using the shifted large-$N$ method to similar calculations as
it yields very accurate results yet remaining simple and straight forward.

\acknowledgments This research was partially supported by the
Scientific and Technological Research Council of Turkey. One of the
authors wishes to dedicate this work to his son Musbah for his love
and assistance.

\bigskip

\begin{table}[tbp]
\caption{Calculated $K$-shell energies $E_{00}$ in $keV$ for some values of $%
Z$}
\begin{tabular}{llllll}
$Z$ & EWA (Ref.5) & Hypervirial-Pad\'{e} (Ref.8) & Shifted--$N$ (Ref.7) &
Expt (Ref.23) & Present Work \\
\tableline$3$ & $-0.053$ $34$ & $-0.054$ $15$ & $-0.054$ $14$ & $-0.054$ $75$
& $-0.054$ $056$ $87$ \\
$4$ & $-0.105$ & $-0.106$ $34$ & $-0.106$ $34$ & $-0.111$ & $-0.106$ $281$ $%
60$ \\
$5$ & $-0.178$ & $-0.180$ $08$ & $-0.180$ $07$ & $-0.188$ & $-0.180$ $078$ $%
08$ \\
$6$ & $-0.274$ & $-0.276$ $23$ & $-0.276$ $23$ & $-0.284$ & $-0.276$ $306$ $%
26$ \\
$7$ & $-0.393$ & $-0.395$ $42$ & $-0.395$ $41$ & $-0.402$ & $-0.395$ $579$ $%
11$ \\
$8$ & $-0.535$ & $-0.538$ $09$ & $-0.538$ $09$ & $-0.532$ & $-0.538$ $354$
\\
$9$ & $-0.701$ & $-0.704$ $61$ & $-0.704$ $61$ & $-0.685$ & $-0.704$ $983$
\\
$14$ & $-1.897$ & $-1.903$ $20$ & $-1.903$ $20$ & $-1.839$ & $-1.904$ $306$
\\
$19$ & $-3.716$ & $-3.725$ $45$ & $-3.725$ $45$ & $-3.607$ & $-3.727$ $639$
\\
$24$ & $-6.171$ & $-6.182$ $77$ & $-6.182$ $77$ & $-5.989$ & $-6.186$ $408$
\\
$29$ & $-9.268$ & $-9.282$ $12$ & $-9.282$ $13$ & $-8.979$ & $-9.287$ $593$
\\
$34$ & $-13.012$ & $-13.028$ $30$ & $-13.028$ $30$ & $-12.658$ & $-13.035$ $%
977$ \\
$39$ & $-17.407$ & $-17.424$ $82$ & $-17.424$ $82$ & $-17.038$ & $-17.435$ $%
077$ \\
$44$ & $-22.454$ & $-22.474$ $38$ & $-22.474$ $38$ & $-22.117$ & $-22.487$ $%
609$ \\
$49$ & $-28.157$ & $-28.179$ $15$ & $-28.179$ $15$ & $-27.940$ & $-28.195$ $%
740$ \\
$54$ & $-34.517$ & $-34.540$ $92$ & $-34.540$ $92$ & $-34.561$ & $-34.561$ $%
250$ \\
$59$ & $-41.535$ & $-41.566$ $12$ & $-41.561$ $17$ & $-41.991$ & $-41.585$ $%
000$ \\
$64$ & $-49.213$ & $-49.241$ $18$ & $-49.241$ $18$ & $-50.239$ & $-49.270$ $%
154$ \\
$69$ & $-57.553$ & $-57.582$ $03$ & $-57.582$ $03$ & $-59.390$ & $-57.615$ $%
917$ \\
$74$ & $-66.554$ & $-66.584$ $70$ & $-66.584$ $70$ & $-69.525$ & $-66.623$ $%
882$ \\
$79$ & $-76.217$ & $-76.250$ $03$ & $-76.250$ $03$ & $-80.725$ & $-76.294$ $%
897$ \\
$84$ & $-86.544$ & $-86.578$ $78$ & $-86.578$ $78$ & $-93.105$ & $-86.629$ $%
718$%
\end{tabular}
\end{table}

\begin{table}[tbp]
\caption{Calculated $K$-shell energies $E_{01}$ in $keV$ for some values of $%
Z$}
\begin{tabular}{llll}
$Z$ & $E_{01}$ & $Z$ & $E_{01}$ \\
\tableline$9$ & $-0.012$ $158$ & $49$ & $-4.207$ $958$ \\
$14$ & $-0.089$ $499$ & $54$ & $-5.358$ $162$ \\
$19$ & $-0.282$ $475$ & $59$ & $-6.655$ $877$ \\
$24$ & $-0.598$ $417$ & $64$ & $-8.102$ $492$ \\
$29$ & $-1.044$ $023$ & $69$ & $-9.699$ $206$ \\
$34$ & $-1.624$ $349$ & $74$ & $-11.447$ $062$ \\
$39$ & $-2.343$ $224$ & $79$ & $-13.346$ $979$ \\
$44$ & $-3.203$ $631$ & $84$ & $-15.399$ $774$%
\end{tabular}
\end{table}

\begin{table}[tbp]
\caption{Calculated $L$-shell energies $E_{10}$ in $keV$ for some values of $%
Z$}
\begin{tabular}{llllll}
$Z$ & EWA (Ref.5) & Hypervirial-Pad\'{e} (Ref.8) & Shifted-$N$ (Ref.7) &
Expt. (Ref.23) & Present work \\
\tableline$9$ & $-0.018$ & $-0.022$ $06$ & $-0.026$ $30$ & $-0.031$ & $%
-0.042 $ $259$ \\
$14$ & $-0.116$ & $-0.124$ $92$ & $-0.124$ $92$ & $-0.149$ & $-0.130$ $396$
\\
$19$ & $-0.320$ & $-0.335$ $03$ & $-0.335$ $03$ & $-0.377$ & $-0.338$ $344$
\\
$24$ & $-0.644$ & $-0.665$ $54$ & $-0.665$ $54$ & $-0.695$ & $-0.669$ $125$
\\
$29$ & $1.096$ & $-1.124$ $48$ & $-1.124$ $48$ & $-1.096$ & $-1.128$ $848$
\\
$34$ & $-1.692$ & $-1.717$ $35$ & $-1.717$ $35$ & $-1.654$ & $-1.722$ $569$
\\
$39$ & $-2.407$ & $-2.448$ $16$ & $-2.450$ $36$ & $-2.373$ & $-2.454$ $212$
\\
$44$ & $-3.272$ & $-3.319$ $98$ & $-3.322$ $03$ & $-3.224$ & $-3.326$ $856$
\\
$49$ & $-4.281$ & $-4.335$ $27$ & $-4.337$ $19$ & $-4.238$ & $-4.342$ $964$
\\
$54$ & $-5.435$ & $-5.496$ $02$ & $-5.497$ $83$ & $-5.453$ & $-5.504$ $555$
\\
$59$ & $-6.737$ & $-6.803$ $90$ & $-6.805$ $63$ & $-6.835$ & $-6.813$ $316$
\\
$64$ & $-8.187$ & $-8.260$ $34$ & $-8.261$ $99$ & $-8.376$ & $-8.270$ $675$
\\
$69$ & $-9.787$ & $-9.866$ $54$ & $-9.868$ $12$ & $-10.116$ & $-9.877$ $861$
\\
$74$ & $-11.538$ & $-11.623$ $58$ & $-11.625$ $10$ & $-12.100$ & $-11.635$ $%
946$ \\
$79$ & $-13.441$ & $-13.532$ $38$ & $-13.533$ $85$ & $-14.353$ & $-13.545$ $%
871$ \\
$84$ & $-15.496$ & $-15.593$ $79$ & $-15.595$ $21$ & $-16.939$ & $-15.608$ $%
473$%
\end{tabular}
\end{table}

\end{document}